\documentclass[allclo]{FBSart}
\usepackage{amsfonts}
\usepackage{amssymb}

\usepackage{graphicx}  
\usepackage{bm}  
\usepackage{amsmath}
\usepackage{epsf}
\newcommand{\beq}{\begin{equation}}
\newcommand{\eeq}{\end{equation}}
\newcommand{\bea}{\begin{eqnarray}}
\newcommand{\eea}{\end{eqnarray}}

\def\vec#1{{\bf #1}}

\title{Universal Properties of the Four-Boson System in Two Dimensions}
\author{L. Platter\instnr{1}\thanks{On leave from FZ J\"ulich,
 Institut f{\"ur} Kernphysik (Theorie), D-52425 J\"ulich
  and HISKP (Theorie), Universit\"at Bonn, Nu\ss allee~14-16, 
  D-53115 Bonn, Germany},
H.-W. Hammer\instnr{1},
Ulf-G. Mei{\ss}ner\instnr{2,3}}
\instlist{Institute for Nuclear Theory, University of Washington, Seattle, 
WA 98195, USA
\and Helmholtz-Institut f\"ur Strahlen- und Kernphysik (Theorie),
Universit\"at Bonn, Nu\ss allee 14-16, D-53115 Bonn, Germany
\and Forschungszentrum J{\"u}lich, Institut f{\"ur} Kernphysik (Theorie),
    D-52425 J{\"u}lich, Germany}
\runningauthor{L.\,Platter, H.\,W.\,Hammer, Ulf-G.\,Mei{\ss}ner}
\runningtitle{Universal Properties of the Four-Boson System in Two Dimensions}
\begin{document}
\maketitle
\begin{abstract}
We consider the nonrelativistic four-boson system in two dimensions
interacting via a short-range attractive potential. For a weakly attractive
potential with one shallow two-body bound state with binding energy 
$B_2$, the binding energies $B_N$ of shallow $N$-body bound states are 
universal and thus do not depend on the details of the interaction potential.
We compute the four-body binding energies in an effective
quantum mechanics approach. There are exactly two bound states: the ground 
state with $B_4^{(0)}=197.3(1)B_2$ and one excited state with 
$B_4^{(1)}=25.5(1)B_2$. We compare our results to recent predictions
for $N$-body bound states with large $N\gg 1$.
\end{abstract}
{\it Introduction}.---Low-dimensional systems are ubiquitous
in many areas of physics. In condensed matter physics, e.g., they are
of interest in connection with high-$T_c$ superconductivity and the 
Quantum Hall effect. A two-dimensional boson system has been realized 
experimentally with hydrogen adsorbed on a helium surface~\cite{Safonov}.
The experimental progress with ultracold atomic gases 
and Bose-Einstein Condensates has made it 
possible to engineer low-dimensional atomic systems in anisotropic traps
\cite{Goerlitz,Schreck,Rychtarik}.

This experimental progress has also stimulated theoretical activities.
Recently, the problem of $N$ weakly attractive bosons in two 
spatial dimensions (2D) was revisited \cite{Hammer:2004as}. 
Relying on the asymptotic freedom of nonrelativistic bosons in 2D with 
attractive short-range interactions, some 
surprising universal properties of self-bound $N$-boson droplets
were predicted for $N\gg 1$.
In particular, the size $R_N$ of the $N$-boson droplets was found to decrease
exponentially with $N$:
\beq
\label{RNratio}
\frac{R_{N}}{R_{N-1}}\approx 0.3417\,,\quad N\gg 1 \,,
\eeq
while their ground state energies $B_N$ increase exponentially with $N$:
\beq
\label{BNratio}
\frac{B_{N}}{B_{N-1}}\approx 8.567 \,, \quad N\gg 1 \,,
\eeq
which implies that the energy required to take out one particle from the 
droplet is about 88\% of the total binding energy. This is in
contrast to typical three-dimensional systems, such as $^4$He droplets, 
where the one-particle separation energy is much less than the total 
binding energy \cite{Pandha}.  
Corrections to Eqs.~(\ref{RNratio}) and (\ref{BNratio}) are expected
to start at order $1/N^2$ and thus should be small for large $N$.
For realistic interactions with a finite range $r_0$, the equations 
are valid for $N$ large, but below a critical value,
\begin{equation}
  1 \ll N < N_{\rm crit} \approx 0.931 \ln\frac{R_2}{r_0}
  + {\cal O}(1)\,.
\label{eq:break}
\end{equation}
At $N=N_{\rm crit}$ the size of the droplet is comparable to the range
of the potential and universality is lost.  If there is a large
separation between $R_2$ and $r_0$, then $N_{\rm crit}$ is much larger
than one.
The universal regime also breaks down when the binding energy of the 
$N$-body states approaches the same order as the particle mass and 
the particles
become relativistic. For typical atomic and molecular systems, however,
the breakdown of the universal regime will be set by the finite range of
the interaction as specified in Eq.~(\ref{eq:break}).

In light of the experimental and theoretical interest in weakly
bound $N$-body clusters in 2D, it is worthwhile to calculate their 
binding energies for short-range interactions explicitly.
For a sufficiently shallow two-body state (such that the 
zero-range approximation can be applied), the binding energies for the 
three-body system in 2D are known exactly. They were first calculated
by Bruch and Tjon in 1979 \cite{Bruch}. More recently, they were
recalculated with higher precision \cite{Nielsen,Hammer:2004as}: 
the ground state has
$B_3^{(0)}=16.522688(1)\, B_2$ and there is one excited state with
$B_3^{(1)}=1.2704091(1)\, B_2$. The value of the ground state energy
$B_3^{(0)}/B_2$ differs from the prediction in Eq.~(\ref{BNratio}) by a factor
of two, indicating that the three-body system is not in the asymptotic regime.
Of course, such deviations from Eq.~(\ref{BNratio}) are expected for 
small values of $N$. However, the exact $N$-body ground state energies should 
approach the prediction (\ref{BNratio}) as $N$ is increased.

{\it The Four-Boson System in 2D}.---
In this paper, we calculate the binding energies of the 
four-body system with attractive, short-range interactions
in 2D. We choose an effective quantum mechanics
approach to generate and renormalize an effective zero-range interaction 
potential that reproduces a given two-body bound state energy.
This potential will then be used in the Yakubovsky equations to compute 
the binding energies of the four-boson system. 
For a sufficiently shallow two-body bound state, the four-body binding 
energies for our effective potential and any realistic finite range 
potential will be the same. In this sense, our results are universal.
This method was recently applied to the four-boson system in three spatial 
dimensions \cite{Platter:2004qn}.
It exploits a separation of scales in physical systems
and is ideally suited to calculate their universal properties.
In principle, finite range corrections to the universal results 
can be calculated systematically within this approach but they 
are beyond the scope of this work.

In two dimensions, any attractive pair potential has at least one
two-body bound state. If this bound state is sufficiently shallow,
$B_2 \ll \hbar^2/(mr_0^2)$, its properties are universal and the 
\lq\lq true'' interaction potential can be replaced by a $\delta$-function
in position space. In momentum space, such a $\delta$-function potential 
corresponds to the interaction
\beq
\label{eq:two-interaction}
V({\bf p},{\bf p'})=\lambda_2\,,\quad \lambda_2 <0\,,
\eeq
which is independent of the relative momenta ${\bf p}$ and ${\bf p'}$
of the incoming and  outgoing pair. 
The bare coupling constant $\lambda_2$ is negative for attractive 
interactions. For convenience, we
work in units where the mass $M$ of the bosons and Planck's constant 
$\hbar$ are set to unity: $M=\hbar=1$.

The interaction (\ref{eq:two-interaction}) is separable 
and the Lippmann-Schwinger equation for the two-body
problem can be solved analytically. The S-wave projected
two-body t-matrix $t(E)$ is given by
\beq
\label{eq:t2}
t(E)=2\pi\Bigr[1/\lambda_2-2\pi\int_0^{\Lambda}\frac{\hbox{d}q
\:q}{E-q^2}\Bigr]^{-1}\,,
\eeq
where $\Lambda$ is an ultraviolet cutoff used to regulate the 
logarithmically divergent integral. The bare coupling
$\lambda_2 = \lambda_2 (\Lambda)$ depends on the cutoff $\Lambda$
in such a way that all low-energy observables are independent of 
$\Lambda$. Demanding that the t-matrix (\ref{eq:t2}) 
has a pole at the bound state energy $E=-B_2$, we can
trade the coupling constant $\lambda_2$ for $B_2$ and obtain
the renormalized t-matrix
\beq
\label{eq:tau}
t(E)=2\Bigl[\log\left(-\frac{B_2}{E}\right)\Bigr]^{-1}\,,
\eeq
which will be used in the following.

The four-body binding energies can be computed by solving the 
Yakubovsky equations \cite{Yakubovsky:1966ue}, which are based on
a generalization of the Faddeev equations for the three-body system. 
The full four-body wave function is first decomposed into Faddeev components, 
followed by a second decomposition into Yakubovsky components. 
In the case of identical bosons, one ends up with two Yakubovsky components 
$\psi_A$ and $\psi_B$. We start from the Yakubovsky equations in the form 
given in \cite{Glockle:1993vr}:
\bea
\psi_A&=&G_0 t_{12} P \Bigl[(1+P_{34})\psi_A + \psi_B\Bigr]\,,
\nonumber\\
\psi_B&=&G_0 t_{12} \tilde{P} \Bigl[(1+P_{34})\psi_A + \psi_B\Bigr]\,,
\label{eq:yak_psi}
\eea
where $G_0=[E-E_{kin}]^{-1}$ is the free four-particle propagator
and $t_{12}$ is the two-body t-matrix in the subsystem of particles 1 and 2.
The operator $P_{ij}$ exchanges particles $i$ and $j$, while $P$ and
$\tilde{P}$ are given by
\beq
P=P_{13}P_{23}+P_{12}P_{23}\,,\qquad\hbox{and}\qquad
\tilde{P}=P_{13}P_{24}\,.
\eeq
Including only S-waves and following the same steps as in 
Ref.~\cite{Platter:2004qn}, we derive a momentum space representation of 
Eqs.~(\ref{eq:yak_psi}) in two spatial dimensions. 
We obtain a system of two coupled integral equations in two variables:
\bea
F_A(u_2,u_3)&=& 2 t(E-{\textstyle\frac{3}{4}u_2^2-\frac{2}{3}
u_3^2})\int_0^\infty\hbox{d}u'_2u'_2 \,
\int_0^{2\pi}\frac{\hbox{d}\phi}{2\pi}\quad
\Bigl[G_0(\hat{u}_2, u'_2, u_3)\,F_A(u'_2,u_3)
\nonumber\\
&&\quad+\int_0^{2\pi}\frac{\hbox{d}\phi'}{2\pi}\:\bigl\{
G_0(\hat{u}_2,\tilde{u}_2, \tilde{u}_3)\,
F_A(\tilde{u}_2, \tilde{u}_3) + G_0(\hat{u}_2,\tilde{v}_2, \tilde{v}_3 )\,
F_B(\tilde{v}_2, \tilde{v}_3)\bigr\}\Bigr]\,,
\label{eq:yaku1}\\[4pt]
F_B(v_2, v_3)&=& 2t(E-{\textstyle\frac{1}{2}v_2^2-v_3^2})
\int_0^\infty\hbox{d}v'_1 v'_1 
\quad \Bigl[ {\textstyle\frac{1}{2}}\, G_0(v_3, v_2, v'_1)\,F_B(v_2, v'_1)
\nonumber\\ &&\quad 
+ \int_0^{2\pi}\frac{\hbox{d}\phi'}{2\pi}\:
G_0( v_3, \bar{u}_2, \bar{u}_3)\,
F_A(\bar{u}_2,\bar{u}_3)\Bigr]\,,
\label{eq:yaku2}
\eea
where we have used the abbreviations
\bea
G_0({u}_1,{u}_2,{u}_3)&=&\left[E-u_1^2-\textstyle{\frac{3}{4}}u_2^2
-\textstyle{\frac{2}{3}}u_3^2\right]^{-1} \,,
\nonumber \\
G_0({v}_1,{v}_2,{v}_3)&=&\left[E-v_1^2-\textstyle{\frac{1}{2}}v_2^2
-v_3^2\right]^{-1}\,,
\eea
and
\bea
\tilde{u}_2 &=&\sqrt{\textstyle{\frac{1}{9}}u_2^2+
\textstyle{\frac{64}{81}}u_3^2
+\textstyle{\frac{16}{27}}u_2 u_3\cos\phi'} \,,\quad
\tilde{u}_3 =\sqrt{u_2^2+
\textstyle{\frac{1}{9}}u_3^2-\textstyle{\frac{2}{3}}
u_2u_3\cos\phi'}\,,
\nonumber \\[0.3cm]
\tilde{v}_2 &=&\sqrt{u_2^2+\textstyle{\frac{4}{9}}u_3^2
+\textstyle{\frac{4}{3}}u_2 u_3 \cos\phi'}
\,,\quad
\tilde{v}_3 =\sqrt{\textstyle{\frac{1}{4}}u_2^2
+\textstyle{\frac{4}{9}u_3^2}
-\textstyle{\frac{2}{3}}u_2 u_3 \cos\phi'}\,,
\nonumber \\[0.3cm]
\bar{u}_2 &=&\sqrt{\textstyle{\frac{4}{9}}v_2^2
+\textstyle{\frac{4}{9}}v_3^2-\textstyle{\frac{8}{9}}
v_2 v_3 \cos\phi'}
\,,\quad
\bar{u}_3 =\sqrt{\textstyle{\frac{1}{4}}v_2^2+v_3^2+
v_2 v_3 \cos\phi'}\,,\nonumber\\[0.3cm]
\hat{u}_2 &=&\sqrt{{\textstyle\frac{1}{4}{u'_2}^2+{u_2}^2+u_2 u'_2
    \cos\phi}}\,.
\eea
The integral equations (\ref{eq:yaku1}) and (\ref{eq:yaku2}) are well
behaved and can be solved numerically without introducing a regulator.

{\it Results}.---The four-body binding energies can be found be discretizing
Eqs.~(\ref{eq:yaku1}, \ref{eq:yaku2}) and calculating the eigenvalues of
the resulting matrix. The energies are given by the energies $E<0$ for
which the matrix has unit eigenvalues while the wave function is given
by the corresponding eigenvectors. Similar to the three-body system, we find
exactly two bound states: the ground state with $B_4^{(0)}=197.3(1)\,B_2$ 
and one excited state with $B_4^{(1)}=25.5(1)\,B_2$.
These binding energies are universal properties of any four-body system in
two spatial dimensions with an attractive short-range
interaction and a sufficiently shallow two-body bound 
state with binding energy $B_2$. Corrections from finite range effects
and three- or four-body forces are expected to be suppressed by powers of 
$r_0^2 E$ where $E$ is the typical energy scale of the problem.
Therefore, finite range corrections will be more important for the
deeper states.

The binding energies for the four-body system in 2D have previously been 
studied for finite-range potentials 
using the Yakubovsky equations \cite{Tjon:1980} as well as 
variational and Monte Carlo methods \cite{Kim:1980,Vranjes}.
In these calculations the two-body binding energy was varied by about
a factor of ten. Over the whole range of binding energies, the ratio
$B_4^{(0)}/B_3^{(0)}$ was found to be approximately 2.9, in apparent
disagreement with our universal result. 
However, one has to keep in mind that the results from these explicit finite
range potentials are far from the universal limit. The two-body
binding energy in these calculations was not small enough for the 
zero-range approximation to be applicable.
From Table III of Ref.~\cite{Tjon:1980}, e.g., it is clear that the 
ratio $B_3^{(0)}/B_2$ is still a factor two away from the universal
zero-range result even for the smallest value of $B_2$ considered.
Furthermore, Fig.~2 of the same reference shows that the dependence
of $B_3^{(0)}/B_2$ on $B_2$ becomes very rapid near $B_2 = 0$. 
Because the four-body states are more deeply bound, it is natural to expect 
an even more rapid dependence of $B_4^{(0)}/B_2$ near
$B_2 = 0$. As a consequence, it will be difficult to obtain the universal 
ratio $B_4^{(0)}/B_3^{(0)}\approx 11.94$ in calculations with finite range 
potentials.

\begin{figure}[tb]
\centerline{\includegraphics*[width=5.in,angle=0]{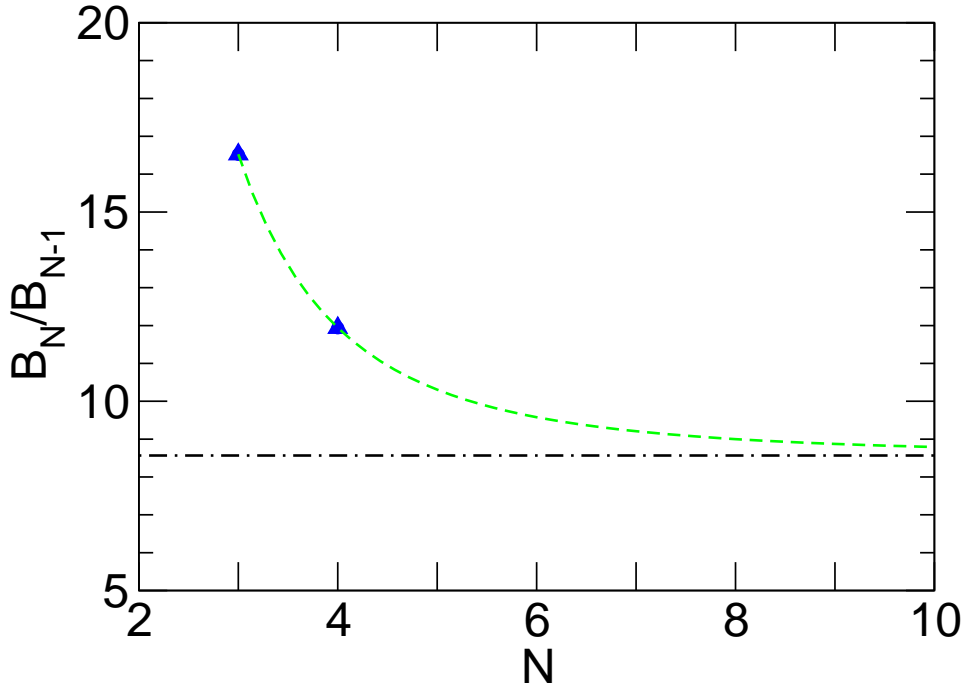}}
\caption{\label{fig:bind}
$B_N/B_{N-1}$ as a function of $N$. The dot-dashed line shows the
asymptotic value of $8.567$. The dashed line is an estimate of how
the large-$N$ value is approached.
}
\end{figure}

We now compare our results with the large-$N$ prediction $B_N/B_{N-1}
\approx 8.567$ from Ref.~\cite{Hammer:2004as}. Using the three-body results
mentioned above, we find $B_4^{(0)}/B_3^{(0)}=11.94$.
This number is considerably closer to the asymptotic value for $B_N/B_{N-1}$ 
than the value for $N=3$: $B_3^{(0)}/B_2=16.52$. 
The exact few-body results for $N=3,4$ are compared
to the asymptotic prediction for $B_N/B_{N-1}$  
indicated by the dot-dashed line in Fig.~\ref{fig:bind}.
The dashed line gives an estimate of how the 
large-$N$ value should be approached. This estimate assumes
the expansion:
\beq
B_N = \left(c_0 +\frac{c_{-1}}{N}+\frac{c_{-2}}{N^2}
+\ldots\right)\, 8.567^N \,,
\label{eq:BNexp}
\eeq
leading to 
\beq
\frac{B_N}{B_{N-1}}= 8.567 +{\cal O}\left(N^{-2}\right)\,.
\label{eq:BNratioexp}
\eeq
The dashed line in  Fig.~\ref{fig:bind} was obtained by fitting the
coefficients of the $1/N$ and $1/N^2$ terms in Eq.~(\ref{eq:BNratioexp})
to the data points for $N=3,4$.
A scenario where the corrections to Eq.~(\ref{eq:BNratioexp}) already start
at ${\cal O}(N^{-1})$ is disfavored by the data points in Fig.~\ref{fig:bind}.

{\it Summary \& Conclusion}.---We have 
computed the four-body binding energies of weakly attractive bosons
with a shallow two-body bound state in 2D. We found exactly two 
bound states and obtained the values $B_4^{(0)}=197.3 (1) \,B_2$ for the 
ground state and $B_4^{(1)}=25.5 (1) \, B_2$ for the excited state. 
The four-body system is considerably closer to the universal 
large-$N$ result ${B_N}/{B_{N-1}}=8.567$ derived in Ref.~\cite{Hammer:2004as}
than the three-body system. 

It would be important to test this prediction at even larger $N$
both theoretically and experimentally. Exact solutions of the 
quantum-mechanical $N$-body equations, however, are currently not 
feasible for $N\geq 5$.
It would be interesting to see whether Monte Carlo methods such as
the combined Monte Carlo/hyperspherical approach used for 3D $^4$He 
clusters in Ref.~\cite{Blume:2000} can be adapted to this problem.
This will not be an easy task, since the calculations for generic
finite range potentials \cite{Tjon:1980,Kim:1980,Vranjes} are not
close to the universal zero-range limit.
Finally, it would be valuable to create bosonic $N$-body clusters
in experiments and measure their size and binding energy. The possibility
of creating such clusters in anisotropic atom traps should be investigated
in the future.

\begin{acknowledge}
We thank D. Blume, D.T.~Son,  and A.~Nogga for discussions.
This work was supported by the German Academic Exchange Service 
(DAAD) and the U.S. department of energy under grant DE-FG02-00ER41132.
\end{acknowledge}

\end{document}